\documentstyle[aps,twocolumn,amssymb,epsf]{revtex}
\def\ave#1{\langle #1\rangle}
\newcommand{\op}[1]{{\hat #1}}

\newcommand{\ma}[1]{{\rm\bf #1}}

\newcommand{\bra}[1]{\langle #1|}
\newcommand{\ket}[1]{|#1\rangle}

\newcommand{\half}{\textstyle{\frac{1}{2}}}
\newcommand{\ad}{{\rm ad\,}}
\newcommand{\tr}{{\rm tr}}
\begin{document}
\title{Quantum invariants of motion in a generic many-body system}
\author{Toma\v z Prosen}
\address{Physics Department, Faculty of Mathematics and Physics,
University of Ljubljana, Jadranska 19, 1111 Ljubljana, Slovenia
}  
\date{\today}
\draft
\maketitle
\begin{abstract}
Dynamical Lie-algebraic method for the construction of local 
quantum invariants of motion in non-integrable many-body systems 
is proposed and applied to a simple but generic toy model, namely an 
infinite kicked $t-V$ chain of spinless fermions. Transition 
from {\em integrable} via {pseudo-integrable (\em intermediate}) 
to {\em quantum ergodic (quantum mixing)} regime in parameter space 
is investigated. {\em Dynamical phase transition} between 
{\em ergodic} and {\em intermediate} (neither ergodic nor completely 
integrable) regime in thermodynamic limit is proposed.
Existence or non-existence of local conservation laws 
corresponds to intermediate or ergodic regime, 
respectively. The computation of time-correlation functions 
of typical observables by means of local conservation laws 
is found fully consistent with direct calculations on finite systems.
\end{abstract}

\pacs{PACS numbers: 03.65.Fd, 05.30.Fk, 05.45.+b}

We investigate the existence of Local Quantum Invariants of motion 
(LQI) (i.e. conservation laws) in a generic quantum many-body system 
of interacting particles. For sufficiently strong non-linear 
coupling between particles one expects quantum
mixing and quantum ergodicity \cite{JLP96} which is incompatible
with existence of LQI \cite{Prelovsek2}.
In this regime (hereafter referred to as {\em quantum ergodic}),
time-correlations between arbitrary pair of quantum observables decay
to zero. This property of quantum ergodicity is
absolutely necessary for the derivation of the laws of
normal transport within linear response theory.
In another extreme case of completely integrable quantum
many-body systems, an infinite number of LQI exist, and
such systems are manifestly non-ergodic.
Anomalous (ideal) transport properties of completely
integrable quantum many body systems have been
discussed in \cite{Prelovsek}.
Our hypothesis is that generic non-integrable quantum
many-body system of (locally) interacting particles may not be 
quantum ergodic, not even in thermodynamic limit (TL) 
(size $\rightarrow\infty$ and fixed density), provided that the 
system is sufficiently close to some completely integrable point 
in the parameter space (analogy with order to chaos transition 
of classical systems of few interacting particles). 
In a recent work \cite{Prosen1} we conjectured 
and demonstrated the existence of {\em intermediate}, 
neither completely integrable nor ergodic, dynamical regime of a 
generic non-integrable quantum many-body system in TL, by directly 
inspecting its time evolution. 
It is the purpose of this paper to show that such an intermediate 
regime may be characterized by {\em quantum pseudo-integrability}: 
existence of at least one or few LQI, which is a sufficient 
condition, using a relation proposed by Mazur \cite{Mazur} and Suzuki 
\cite{Suzuki}, to prevent time correlations of certain observables 
to decay to zero.

In \cite{Prosen1} a novel family of simple but generic many-body 
systems of locally interacting particles has been introduced 
smoothly interpolating between integrable and ergodic regime, 
namely {\em kicked t-V model} (KtV) of spinless fermions with 
periodically switched nearest neighbor interaction on an
infinite chain with time-dependent Hamiltonian
\begin{equation}
H(\tau) = \sum_j\left[-\half t(c^\dagger_j c_{j+1} + h.c.) +
\delta_p(\tau) V n_j n_{j+1}\right],
\label{eq:hamiltonian}
\end{equation}
$c_j^\dagger,c_j,n_j=c^\dagger_j c_j$ are fermionic creation, 
annihilation and number operators, respectively, and 
$\delta_p(\tau)=\sum_{m=-\infty}^\infty \delta(\tau-m)$.
The time-dependent Hamiltonian (\ref{eq:hamiltonian})
can be written as $H(\tau) = t H_1 + \delta_p(\tau) V H_0$ where
the dimensionless kinetic energy $H_1$ and the kick potential $H_0$ 
may be rewritten in terms of independent spin variables 
$(\sigma^{\pm}_j,\sigma^z_j)$ on sites $j$, via Wigner-Jordan 
transformation,
$$
H_1 = 
\half \sum_j (\sigma^+_j \sigma^-_{j+1} + \sigma^-_j \sigma^+_{j+1}),
\quad
H_0 = 
\half \sum_j \sigma^z_j \sigma^z_{j+1}.$$
Symmetric time evolution of KtV system for one period is given by 
explicit unitary quantum many-body map ($\hbar=1$),
$U = \hat{T}\exp(-i\int_{-1/2}^{1/2} d\tau H(\tau)) = 
\exp(-i t H_1/2)\exp(-i V H_0)\exp(-i t H_1/2).$ 
Note that $V$ is a cyclic parameter of period $2\pi$, and the dynamics
is essentially invariant w.r.t. transformations 
$t\rightarrow -t$ and $V\rightarrow -V$, so we consider only the 
half-strip $(t,V)\in[0,\infty)\times[0,\pi)$.
KtV model is completely integrable for: $t=0$ (Ising model), 
or $V=0\pmod{2\pi}$ (free fermion model), or  
$t,V\rightarrow 0$ and $t/V$ finite (XXZ model).

We prefer to consider Heiseberg representation and
write a map over the algebra $\frak{U}$ of quantum 
observables $A(\tau)$ for time evolution over one period,
$\op{U}_{\ad} : A(n-\half) \rightarrow A(n+\half)
= U^\dagger A(n-\half) U$, explicitly as
$$
\op{U}_{\ad} = \exp\left(\frac{it}{2} \ad H_1\right) 
               \exp\left(iV \ad H_0\right)
               \exp\left(\frac{it}{2} \ad H_1\right).
$$
where $\ad$ is the usual adjoint map on the Lie algebra $\frak{U}$,
$\ad A : B \rightarrow [A,B] = AB-BA$.
Infinite-dimensional Lie algebra $\frak{U}$ of
local quantum observables is also a {\em Hilbert space} equiped with the 
{\em invariant bilinear form}
\begin{equation}
(A|B) = \lim_{L\rightarrow\infty} \frac{1}{L 2^L}
\tr_L A^\dagger B
\label{eq:metric}
\end{equation}
which is a scalar product invariant under the adjoint map 
$\ad A:B\rightarrow[A,B]$,
$((\ad A^\dagger)B|C) = (B|(\ad A)C)$.
The limit and the trace $\tr_L$ is defined thru finite systems
of increasing size $L$.
Our aim is to check the existence of LQI which are the 
(normalizable \cite{Locality}) fixed points $A$ ($(A|A) < \infty$) 
of the Heissenberg map $\op{U}_\ad$
\begin{equation}
\op{U}_{\ad} A = A.
\label{eq:fixp}
\end{equation}
However, we do not suggest to solve eq. (\ref{eq:fixp}) in the full
algebra $\frak{U}$ which is a highly prohibitive task.
Instead, we devise a special subalgebra of $\frak{U}$,
the Minimal Invariant Lie Algebra (MILA), which is invariant
to motions generated by the kinetic or the potential part of the
Hamiltonian and hence it is also invariant to $U_{\ad}$. 
Having the two generators, $H_0$ and $H_1$, spanning 2-dim
subspace ${\frak s} = \{\alpha H_0 + \beta H_1\}$,
we construct the basis of MILA ${\frak S}=\bigcup_{n=1}^\infty 
(\ad {\frak s})^n {\frak s}$ ordered by the order of locality
as follows: We assign an observable 
$\tilde{H}_{p,b}$ to an {\em ordered pair} of integers $(p,b)$, 
{\em order} $p$, and {\em code} $b,\, 0\le b < 2^p$ with $p$
binary digits $b_n,\,b=\sum_{n=0}^{p-1} b_n 2^n$, namely
$$\tilde{H}_{p,b} = 
(\ad H_{b_{p-1}})(\ad H_{b_{p-2}})\cdots(\ad H_{b_1}) H_{b_0}.$$
Since not all observables $\tilde{H}_{p,b}$ upto a given maximal 
order $q, p\le q,$ are linearly independent we perform Gram-Schmit 
orthogonalization w.r.t. the scalar product (\ref{eq:metric})
\begin{eqnarray}
G_{q,c} &=& \cases{
\tilde{G}_{q,c}/\sqrt{(\tilde{G}_{q,c} | \tilde{G}_{q,c})}; 
& $\tilde{G}_{q,c} \neq 0$,\cr
0; & $\tilde{G}_{q,c} = 0$,\cr }
\nonumber \\
\tilde{G}_{q,c} &=& \tilde{H}_{q,c} - 
\sum_{(p,b)}^{(p,b)<(q,c)}
G_{p,b} (G_{p,b}|\tilde{H}_{q,c}).
\label{eq:base}
\end{eqnarray}
The nonzero (normalized) local observables 
$G_{q,c}$ form the orthonormal basis of
MILA. Note that observables $G_{q,c}$ are local
operators of order $q$, they have been expanded in terms
of spatially homogeneous finite products of field operators, 
say 
$Z_{s_0 s_1\ldots s_q}=\sum_j \sigma_j^{s_0}\sigma_{j+1}^{s_1}\cdots
\sigma_{j+q}^{s_q}$,
where $s_k\in\{0,+,-,z\}$ and $\sigma_j^0=1$.
The number of terms $Z_{s_0\ldots s_q}$ in expansion
of $G_{q,c}$ was found to grow exponentially as 
$\sim 2.55^q$ (on average).
The observables $Z_{s_0\ldots s_q}$ form a convenient
orthonormal Euclidean basis w.r.t. (\ref{eq:metric}) 
of the Hilbert space ${\frak H}$ of local spatially 
homogeneous observables.
Let us now consider truncated linear subspaces of
MILA, ${\frak S}_p=\bigcup_{n=0}^p 
(\ad {\frak s})^n {\frak s}$, with dimensions 
$d_p = \dim{\frak S}_p$,
linearly spanned by observables $G_{q,c}$ up to maximal order $p$, 
$q\le p$.
Let $\ma{H}_{p,\alpha}$ denote real and symmetric (Hermitean in general)
matrices of linar maps $\ad H_\alpha$ on ${\frak S}_p$ with
images orthogonally projected back to ${\frak S}_p$.
It follows from the construction that they have (generally) 
a block-band structure where blocks correspond to
observables with fixed order $q$: namely
$(G_{q,c}|\ad H_\alpha|G_{q',c'})\neq 0$ only if 
$|q-q'|=1$.

Our program is to solve the equation (\ref{eq:fixp}) 
numerically in truncated subspaces of MILA, ${\frak S}_p$,
and check whether the procedure converges as $p$ increases.
This is quite feasible since the dimensions
$d_p=\dim{\frak S}_p$ increase much less rapidly than, say, 
the dimensions of truncated subspaces  
of a huge Lie algebra ${\frak H}$ of homogeneous observables, 
spanned by $Z_{s_0\ldots s_p}$. 
In case of KtV the former increase
approximately as $d_p \approx 1.68^{p-1}$ 
(see table \ref{tab:1})
while the latter goes as $\sim 4^{p+1}$.  
The truncated adjoint maps, $\ma{H}_{p,\alpha}$, 
have nontrivial null spaces ${\frak N}_{p,\alpha} = 
\{ A\in {\frak S}_p,[H_\alpha,A]\in 
{\frak S}_{p+1}-{\frak S}_p\}$, 
with dimensions $d_{p,\alpha}=\dim{\frak N}_{p,\alpha}$ which
increase approximately with the same exponent $\propto 1.68^p$.
By means of computer algebra we managed to go as high as $p=14$. 
An important observation was that a matrix
$1 - \exp(i\half t \ma{H}_{p,1})
     \exp(iV \ma{H}_{p,0})
     \exp(i\half t \ma{H}_{p,1})$ possesses a high-dimensional 
null space ${\frak N}_p(t,V)$ whose dimension is, for odd $p$, 
{\em independent} of parameters $t,V$ and equal to the 
dimension of nullspace of $\ma{H}_{p,1}$, $\dim{\frak N}_{2l-1} = 
d_{2l-1,1}.$ 
Note also that for odd order of truncation $p=2l-1$, the elements of 
null space $A\in {\frak N}_p(t,V)$ are spanned by combinations
of {\em odd} powers of generators, i.e. 
$(A|G_{2l,c})\equiv 0$, which is due to
time-symmetric construction of evolution operator $\op{U}_{\ad}$.
However, we are seeking for LQI $A\in {\frak S}$, which are normalizable,
i.e. the {\em relative norm} in the subspace of local operators of 
fixed order $p$, defined as $N_q(A) := \sum_c |(A|G_{q,c})|^2$, should 
be a rapidly decreasing function of the order $q$, since 
$(A|A) = \sum_{q=1}^{\infty} N_q(A) < \infty$.
Only the elements of ${\frak N}_p(t,V)$ which are within certain
numerical accuracy independent of the order of truncation $p$ are
candidates for LQI.
To find them we minimize a quadratic form,
the relative norm at truncation order $N_p(A)$,
i.e. we diagonalize the operator 
$\op{N}_p = \sum_c G_{p,c} \otimes G_{p,c}$ 
in the subspace ${\frak N}_p(t,V)$.
In Fig.1 we plot the relative norm $N_q(A_m)$ of the first three 
eigenvectors $A_m$ 
of $\op{N}_p$ with $p=13$ corresponding to the smallest values of 
quadratic form $N_p(A_m)$ againts the (odd) order 
$q \in\{1,3,\ldots,p\}$. We give a mesh of plots
for various values of the parameters $t,V$. Note that 
$N_{2l}(A_m)\equiv0$ due to symmetry.
In certain region of parameter space $t,V$ we have found a good 
agreement with {\em exponential} 
\begin{equation}
N_{2l-1}(A_m) \propto \exp(-s_m l)
\label{eq:expon}
\end{equation} 
with a positive exponent $s_m(t,V)$ for the first observable 
$m=1$, and in a smaller subregion of parameter space even for the 
second observable $m=2$, however with a smaller exponent $s_2$.
Positivity of $s_m$ as determined from exponential fit (\ref{eq:expon}) 
for $2l-1=3,5\ldots p-2$ has been used as a numerical creterion
of convergence (locality) of $A_m$.
By comparing with results for $p=11$ we have checked that
the converged observables are (almost) 
independent on the variation of the truncation order $p$. 
For the rest of the null space ${\frak N}_p(t,V)$, $m>2$,
we found roughly a uniform distribution $N_{2l-1}(A_m)\sim 1$ so 
these observables cannot converge as $p\rightarrow\infty$.
We conclude that this is an evidence for the existence of
one or two LQI in certain region in the
parameter space, roughly where $t < 1.5$.
Outside this region, all exponents $s_m(t,V)$ are zero 
and none of the eigenvectors $A_m$ converges to a LQI which 
is consistent with the property of quantum ergodicity.
In Fig.2 we show a phase-diagram of an exponent 
$s_1(t,V)$ of LQI $A_1(t,V)$. In table \ref{tab:2} we give 
explicitly the first few coefficients $f_{p,c}$, of expansion 
$A = \sum_{(p,c)} f_{p,c} G_{p,c}$ of the one LQI for $t=V=1$, 
and of the two LQI for $t=0.2,V=1$\cite{Divergence}.

\begin{figure}[htbp]
\begin{center}
{\leavevmode
\epsfxsize=3truein
\epsfbox{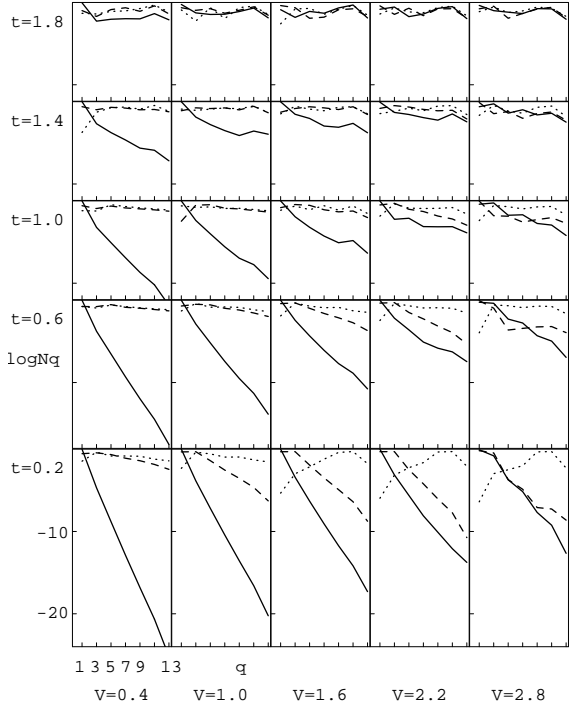}}
\end{center}
\caption{The logarithmic relative norm of the first three candidates 
$A_m,m=1,2,3$ for LQI, $\log_{10}N_q(A_m)$, 
is plotted against (odd) order $q=2l-1$,
for a square mesh of parameters $t$ and $V$. 
The order of truncation is $p=13$. Full curves: $m=1$, dashed curves: $m=2$, and
dotted curves: $m=3$.
}
\label{fig:1}
\end{figure}
\begin{figure}[htbp]
\begin{center}
\vspace{-6mm}
{\leavevmode
\epsfxsize=3.2truein
\epsfbox{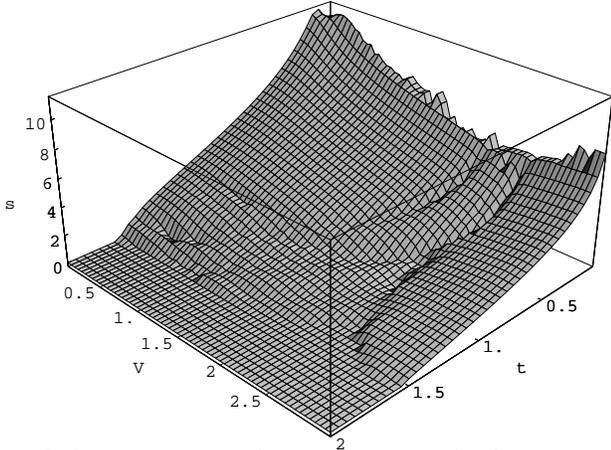}}
\vspace{-4mm}
\end{center}
\caption{
The largest inverse order-localization length $s_1$ for the first 
numerical LQI $A_1$ ($p=13$) vs. parameters $t,V$.}
\label{fig:2}
\end{figure}
\vspace{-4mm}
\begin{table}
\begin{tabular}{|c|ccccccccccccc|}
$p$    
&  2 &  3 &  4 &  5 &  6 &  7 &  8 &  9 &  10 &  11 &  12 &  13 &  14
\\ \hline  
$d_p$ 
& 3 & 5 & 7 & 11 & 16 & 26 & 41 & 67 & 108 & 179 & 294 & 495 & 832 
\\ \hline
$d_{p,0}$
& 1 & 2 & 3 &  5 &  6 & 10 & 13 & 23 & 34 & 61 & 92 & 163 & 258
\\ \hline
$d_{p,1}$
& 1 & 2 & 1 &  5 &  2 & 10 &  7 & 21 & 22 & 51 & 66 & 137 & 202
\end{tabular}
\caption{Dimensions of truncated MILA ${\frak S}_p$, and of nullspaces
of $\ma{H}_{p,\alpha}$, for different orders of truncation $p$.}
\label{tab:1}
\end{table}

Further, we use a theorem of Mazur \cite{Mazur} generalized by Suzuki 
\cite{Suzuki} (MS) to compute canonical averages 
\cite{Average} of time averaged correlation functions of certain 
observables. We consider dimensionless kinetic energy 
$T = H_1$, where $\ave{T}=(1|T) = 1$, with time-correlator 
$$ D = 
\lim_{N\rightarrow\infty}\frac{1}{N}
\sum_{n=0}^N ([T(\half)-\ave{T}]|[T(n+\half)-\ave{T}]).
$$ 
MS equation expresses $D$ in terms of a sum over all LQI   
\begin{equation}
D = \sum_m |(A_m|T)|^2/(A_m|A_m).
\label{eq:MS}
\end{equation}
In case of complete integrability there are {\em infinitely} many LQI,
in case of quantum ergodicity there are {\em none} and MS correlator $D$ is
zero, while in intermediate regime
the sum (\ref{eq:MS}) (for KtV) has one or two nonzero terms.
In Fig.3 we show a phase diagram of the kinetic time-correlator $D$ as
determined from LQI (\ref{eq:MS}).
Note a sharp (phase) transition between ergodic dynamics 
(disordered phase $D=0$)
and intermediate dynamics (ordered phase $D>0$) which
may also be characterized by the maximal inverse 
{\em order-localization length}
$s_1$ (Fig.2) which linearly decreases to zero at the transition. 
We have compared the above results on infinite KtV chains to direct 
calculations on finite chains of $L$ sites with periodic 
boundary conditions, $c_{L+1} = c_1$, which have a discrete quasi-energy 
spectrum $\eta_n$ and eigenstates $\ket{n}$, 
$U\ket{n}=\exp(-i\eta_n)\ket{n}$. For a finite system of size $L$, 
a time-correlator reads
\begin{equation}
D_L = 2^{-L} \sum_{n=1}^{2^L} ({\textstyle{\frac{1}{L}}}\bra{n}T\ket{n} 
- \ave{T})^2.
\label{eq:DL}
\end{equation}
However, $D_L$ need not necessarily converge to the proper
time-correlator of an infinite system $L=\infty$, since then
the time-limit $\tau\rightarrow\infty$ 
is taken prior to the size-limit $L\rightarrow\infty$ \cite{JLP96}.
Nevertheless the behavior of $D_L$ for $L=16$
shown in Fig.4 as a function of parameters $t,V$ is very similar 
to the infinite system MS correlator (\ref{eq:MS}) shown in Fig.3. 
Agreement is even quantitative, except in the region of 
transition between dynamical phases (Fig.5). 

\begin{table}
\begin{tabular}{|r||r|r|r|}
term & $A_1(t=0.2)$ & $A_2(t=0.2)$ & $A_1(t=1)$ \\ \hline
$G_{1,0}$     & -.9652693 & -.18626 & -.61162 \\
$G_{1,1}$     & -.2609845 & 0.65767 & -.78843 \\
$G_{3,001}$   & 0.0116246 & -.69908 & 0.04195 \\
$G_{3,101}$   & 0.0026499 & -.02756 & 0.04928 \\
$G_{5,01001}$ & $2.487\cdot 10^{-4}$ & 0.20008 & 0.00507 \\
$G_{5,01101}$ & $4.200\cdot 10^{-5}$ & 0.00337 & 0.00450 \\
$G_{5,11001}$ & $4.533\cdot 10^{-5}$ & 0.00865 & 0.00487 \\
$G_{5,11101}$ & $9.884\cdot 10^{-6}$ & 0.00056 & 0.00555 
\end{tabular}
\caption{Coefficients of expansion of the two LQI (col.2,3)
for $t=0.2,V=1$ and of the single LQI (col.4) for $t=V=1$ in 
terms of basis $G_{q,c}$ (col.1, $c$ is a binary code) up to 
6th order. Truncation order is $p=13$. Note that
all the digits shown (except possibly the last one) 
are the same for truncation at $p=11$. 
Other obsevables $G_{q,c}, q=1,3,5$, are zero by
construction (\ref{eq:base}).}
\label{tab:2}
\end{table}

Note that KtV map $\op{U}_\ad$ is invariant under the parity operation 
$\op{P}: c_j\rightarrow c_{-j}$ and MILA may exhaust \cite{Exhaust}
only the positive parity class, $\op{P}A=A$. Unfortunately,
the {\em current} $J=i(\sum_j c_j^\dagger c_{j+1} - h.c.) $, 
which, interestingly, gives rise to ideal transport 
in intermediate \cite{Prosen1} (integrable \cite{Prelovsek2,Prelovsek})
regime, has a negative parity $\op{P}J=-J$, and hence zero overlap with 
the above LQI of MILA, $(A_m|J) = (\op{P}A_m|\op{P}J) = -(A_m|J) = 0$.
However, analogous results have been found for the negative parity 
class of observables as well \cite{Prosen2}.

The total occupation number 
$N = \sum_j n_j = \half \sum_j (\sigma^z_j+1)$,
is a trivial invariant of motion
since $(N|{\frak S}) = (N|J) = 0$.
One may either consider observables over the Fock subspace of states with a fixed 
density $\rho:=\ave{N}\in[0,1]$, or as we do here using a full trace 
average (\ref{eq:metric}), consider the entire Fock 
space, which is in TL equivalent to half filling, $\rho=\half$.

\begin{figure}[htbp]
\begin{center}
\vspace{-5mm}
{\leavevmode
\epsfxsize=3.2truein
\epsfbox{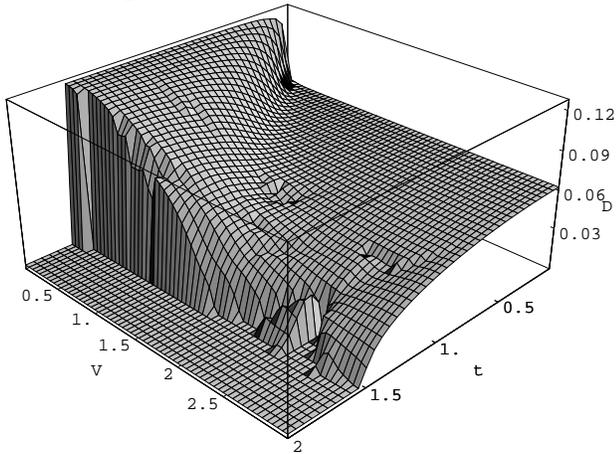}}
\vspace{-3mm}
\end{center}
\caption{
MS correlator (\ref{eq:MS}) $D$ ($p=13$) vs. parameters $t,V$.
}
\label{fig:3}
\end{figure}

\begin{figure}[htbp]
\begin{center}
\vspace{-5mm}
{\leavevmode
\epsfxsize=3.2truein
\epsfbox{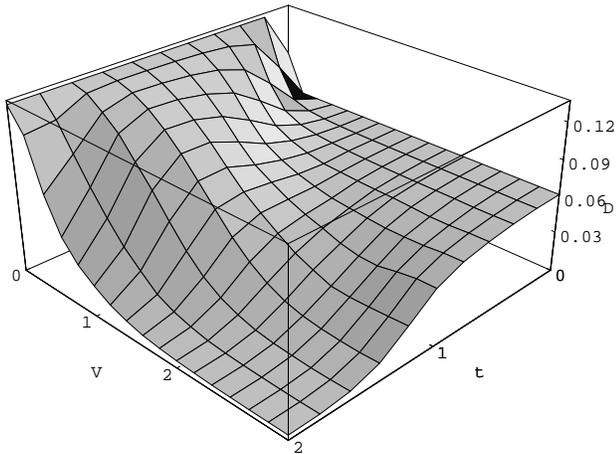}}
\vspace{-3mm}
\end{center}
\caption{
The finite size kinetic correlator (\ref{eq:DL}) $D_L$ for $L=16$
vs. parameters $t,V$.
}
\label{fig:4}
\end{figure}

\noindent
{\em Conclusions:}
We have found strong evidence for the existence of non-trivial LQI, 
in a simple but generic quantum many-body system in TL, namely kicked 
t-V model \cite{Prosen1}. The algebraic method, which should be 
applicable to other non-integrable quantum many-body systems, is based
on the (computerized) construction of minimal invariant
infinitely dimensional Lie algebra, MILA, generated by the 
essential parts of Hamiltonian
(in our case, by kinetic energy and kick potential).
LQI are found numerically as fixed points of the adjoint map of the 
evolution operator (or of the Hamiltonian if system was autonomous) 
in MILA. Existence of LQI is found to be fully consistent with deviations
from quantum ergodicity characterized by non-vanishing averaged 
time-autocorrelations $D$ of a typical observable; here we use 
the kinetic energy. $D$ is a suitable order parameter describing
the phase transition from the pseudo-integrable (intermediate) regime
($D > 0$) to the quantum ergodic regime ($D=0$).

Discussions with Prof. P. Prelov\v sek, and the financial support by the
Ministry of Science and Technology of R Slovenia are 
gratefully acknowledged.

\begin{figure}[htbp]
\begin{center}
{\leavevmode
\epsfxsize=3.2truein
\epsfbox{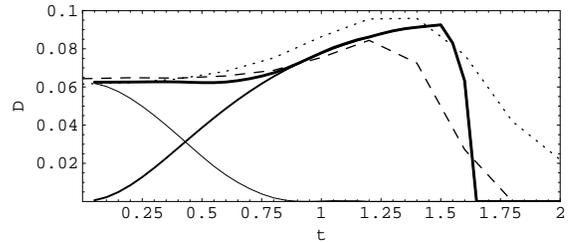}}
\end{center}
\caption{
The comparison between MS correlator (\ref{eq:MS}) $D$
of infinite KtV chain (thick full curve) and
finite size correlator (\ref{eq:DL}) $D_L$ for $L=16$
(dotted curve) and linearly extrapolated to $1/L=0$ from data for $L=16$ and $L=12$
(dashed curve) vs. parameter $t$, and for fixed parameter $V=1$.
Medium full curve and thin full curve denote separate contributions
of the first and second LQI to MS correlator $D$ (\ref{eq:MS}),
respectively.
}
\label{fig:5}
\end{figure}


\begin{references}
\bibitem{Locality} Note that non-local obervables
are not normalizable in the metric (\ref{eq:metric}).
\bibitem{Divergence} Of course, at present we cannot completely 
exclude the possibility that such expansions are generically divergent
asymptotic series.
\bibitem{Average} Our example corresponds to infinite temperature,
since we use a simple trace measure (\ref{eq:metric}).
\bibitem{Exhaust} We have tried as well to search for LQI in 
the entire (huge) algebra of homogeneus observables ${\frak H}$
(truncated at order $p=6$), i.e. we have numerically solved
eq. (\ref{eq:fixp}) for the matrix of the map $U_{\ad}$ in the
basis of observables $Z_{s_0\ldots s_p}$, and we found roughly 
identical (and no more) numerical LQI as in MILA, 
although ${\frak S} \subset 
{\frak H}$.
\end{references}
\end{document}